\pdfoutput=1
\PassOptionsToPackage{bookmarks=false,colorlinks=true,linkcolor=blue,urlcolor=cyan,filecolor=magenta,citecolor=red,pdfstartview=FitV,hyperfootnotes=false,pdftitle={Boundary Carrollian CFTs and Open Null Strings},pdfauthor={Arjun Bagchi, Pronoy Chakraborty, Shankhadeep Chakrabortty, Stefan Fredenhagen, Daniel Grumiller, Priyardarshini Pandit},pdfsubject={open null strings},pdfkeywords={open null strings and boundary carrollian conformal field theories},bookmarksopen=true}{hyperref}
\documentclass[aps,prl,reprint,superscriptaddress]{revtex4-2} 
  
\usepackage[dvipsnames]{xcolor}
\usepackage[bookmarks=false,colorlinks=true,linkcolor=blue,urlcolor=cyan,filecolor=magenta,citecolor=red,pdfstartview=FitV,hyperfootnotes=false,pdftitle={Boundary Carrollian CFTs and Open Null Strings},pdfauthor={Arjun Bagchi, Pronoy Chakraborty, Shankhadeep Chakrabortty, Stefan Fredenhagen, Daniel Grumiller, Priyardarshini Pandit},pdfsubject={open null strings},pdfkeywords={open null strings and boundary carrollian conformal field theories},bookmarksopen=true]{hyperref}
\usepackage{amssymb,amsmath,amsfonts}
\usepackage{epsfig}
\usepackage{graphicx}
\usepackage{epstopdf}
\usepackage{natbib}
\usepackage{braket}
\usepackage{dutchcal}
\usepackage{mathrsfs}

\newcommand{\be}{\begin{eqnarray}}
	\newcommand{\ee}{\end{eqnarray}}
\newcommand{\beq}{\begin{eqnarray}}
	\newcommand{\eeq}{\end{eqnarray}}

\newcommand{\beqa}{\begin{eqnarray}}
	\newcommand{\eeqa}{\end{eqnarray}}

\renewcommand{\L}{\mathcal{L}}

\definecolor{gris}{rgb}{0.5,0.5,0.5}
\definecolor{darkgreen}{rgb}{0.0,0.5,0.0}

\DeclareMathOperator{\extdm}{d}
\newcommand{\extd}{\extdm \!}

\allowdisplaybreaks[0]

\begin{document}

%%%%%%%%%%%%%%%%%
%%% TITLEPAGE %%%
%%%%%%%%%%%%%%%%%	
		
	\title{Boundary Carrollian CFTs and Open Null Strings}
	
\author{Arjun Bagchi}
\email{abagchi@iitk.ac.in}
\affiliation{Indian Institute of Technology Kanpur, Kanpur 208016, India}	
\author{Pronoy Chakraborty} 
\email{pronoy.22phz0009@iitrpr.ac.in}
\affiliation{ Indian Institute of Technology Ropar, Rupnagar, Punjab 140001, India}
\author{Shankhadeep Chakrabortty} 
\email{s.chakrabortty@iitrpr.ac.in}
\affiliation{ Indian Institute of Technology Ropar, Rupnagar, Punjab 140001, India}
\author{Stefan Fredenhagen} 
\email{stefan.fredenhagen@univie.ac.at}
\affiliation{University of Vienna, Faculty of Physics, Boltzmanngasse 5, 1090 Vienna, Austria}
\author{Daniel Grumiller} 
\email{grumil@hep.itp.tuwien.ac.at}
\affiliation{Institute for Theoretical Physics, TU Wien, Wiedner Hauptstrasse 8–10, 1040 Vienna, Austria}
\author{Priyadarshini Pandit}
\email{ppandit@iitk.ac.in}
\affiliation{Indian Institute of Technology Kanpur, Kanpur 208016, India}
	
\begin{abstract}
We consider Carrollian conformal field theories in two dimensions and construct the boundary Carrollian conformal algebra (BCCA), opening up innumerable possibilities for further studies, given the growing relevance of Carrollian symmetries. We prove that the BCCA emerges by contracting a single copy of the Virasoro algebra. As an application, we construct, for the first time, open null strings and show that, for Dirichlet boundary conditions, we recover the BCCA as the algebra of constraints. We finally reconstruct our string results by taking the null limit of tensile open strings.

\end{abstract}
		
\maketitle

%%%%%%%%%%%%%%%%%
%%% WORD COUNT %%
%%%%%%%%%%%%%%%%%

% 38 displayed Eq. lines: 40*16 = 640 words
% 1 Fig. with aspect ratio 1: 170 words
% text 2930 words (29.8., 15:00 Vienna time)
% total 3735 words - we have 15 words available
% allowed maximum in PRL is 3750 words
%
% abstract has an upper bound of 600 characters (including white spaces and punctuation)
% current abstract has 590 characters - we have 10 characters available 

%%%%%%%%%%%%%%%%%
%%% MAIN TEXT %%%
%%%%%%%%%%%%%%%%%
 
\paragraph{Introduction.} 
The study of conformal field theories (CFTs) with boundaries has a rich history and a wide variety of applications to condensed matter systems, e.g., the Kondo problem \cite{Kondo:1964nea,Affleck:1990zd,Affleck:1990iv}, quantum wires \cite{vanWees:1988zz}, or quantum systems with dissipation \cite{Johanesson:2003}. Boundary CFTs (BCFTs) also arise in string theory through open strings. The techniques of BCFTs have been particularly useful for the study of D-branes in string theory \cite{Pradisi:1988xd,Polchinski:1987tu,Polchinski:1995mt,Polchinski:1996na,Karch:2000gx,Recknagel:2013uja}. 

The study of BCFTs in two dimensions (2d) was pioneered by Cardy \cite{Cardy:1984bb}. Here constraints imposed by infinite conformal symmetries lead to numerous exact results for correlation functions, critical exponents, entanglement entropy, and other observables \cite{Cardy:1989ir,Cardy:1991tv,Fateev:2000ik,Calabrese:2004eu,Calabrese:2006rx}. BCFTs have also found interesting applications in holography following the AdS/BCFT proposal \cite{Takayanagi:2011zk}. 

In this Letter, we focus on a different class of CFTs, the so-called Carrollian Conformal Field Theories (CCFTs). The Carroll algebra is obtained as a vanishing speed of light limit of the Poincar\'e algebra \cite{Leblond65,SenGupta:1966qer,Henneaux:1979vn} and has recently attracted a lot of attention due to its emergence in many places of interest. In particular, Carrollian symmetry seems to be central to some condensed matter systems (fractons \cite{Bidussi:2021nmp}, systems with flat bands \cite{Bagchi:2022eui}), in hydrodynamics of the quark-gluon-plasma \cite{Bagchi:2023ysc,Bagchi:2023rwd}, in cosmology \cite{deBoer:2021jej} and on the event horizons of generic black holes \cite{Henneaux:1979vn,Donnay:2019jiz}. Conformal extensions of Carrollian symmetries that underlie CCFTs have been proposed as holographic duals of asymptotically flat spacetimes \cite{Bagchi:2010zz,Barnich:2012aw,Bagchi:2012yk,Bagchi:2012xr,Barnich:2012xq,Bagchi:2012cy,Bagchi:2016bcd,Donnay:2022aba,Bagchi:2022emh,Donnay:2022wvx, Bagchi:2023fbj} and these symmetries have also been discovered on the worldsheet of tensionless null strings \cite{Bagchi:2013bga,Bagchi:2019cay}. 

The current Letter aims to put forward, for the first time, a CCFT with boundaries. We focus on 2d, where the limit from the relativistic CFT is more transparent. As explained below, we discover, to the best of our knowledge, a new symmetry algebra not encountered before in the field (or, for that matter, in other branches of physics). As an example, we show that the same symmetries arise when one considers open null tensionless strings. The construction of these open null strings is also done for the first time in this Letter. 

The tensionless regime of string theory is the opposite of the more explored and better-understood point particle limit, where the fundamental string shrinks to a point and reproduces Einstein gravity and quantum field theory. Tensionless strings are expected to capture the ultrahigh-energy sector of strings and be an indicator of what lies beyond the mysterious Hagedorn phase transition \cite{Atick:1988si, Pisarski:1982cn, Olesen:1985ej}. In flat spacetimes, the tensionless limit is synonymous with the null limit of strings, which is the one-dimensional extended analog of massless point particles. This was first studied by Schild in \cite{Schild:1976vq}, and \cite{Karlhede:1986wb,Lizzi:1986nv,Gamboa:1989zc} are some early investigations in this direction. Of particular importance is the work by Isberg et.~al.~\cite{Isberg:1993av} (ILST). 

The study of tensionless strings was given a new life in \cite{Bagchi:2013bga} following the realization that the symmetries that appear on the worldsheet are equivalent to the symmetries of the 3d Bondi--Metzner--Sachs (BMS$_3$) algebra, which appears in the context of asymptotically flat spacetimes at their null boundary \cite{Bondi:1962px,Sachs:1962wk,Ashtekar:1996cd}. In modern parlance, the conformal Carrollian algebra replaces the two copies of the Virasoro algebra as the worldsheet symmetry of the tensionless string. This is based on an isomorphism between the conformal Carroll and the BMS algebras \cite{Duval:2014uva}. 

Following ILST, the recent literature \cite{Bagchi:2015nca,Casali:2016atr,Bagchi:2020fpr,Bagchi:2020ats,Bagchi:2021rfw,Bagchi:2021ban,Bagchi:2022iqb,Banerjee:2023ekd,Chen:2023esw,Bagchi:2023cfp,Banerjee:2024fbi,Bagchi:2024rje} has focused entirely on the construction of the closed null string. This stemmed from the fact that algebraically, it is well understood how to get to the BMS$_3$ algebra by an In{\"o}n{\"u}--Wigner contraction of two copies of the Virasoro algebra. The na{\"i}ve contraction of a single Virasoro algebra, on the other hand, is trivial, leaving one with a rather reduced algebra from which it is difficult to see how an open tensionless string would emerge. See, however, \cite{Sagnotti:2011jdy, Bonelli:2003kh} for some attempts at this contraction and connections to higher spin theories. 

In this Letter, we take a different route and construct, for the first time, an open null tensionless string. We study the open tensionless string intrinsically from the ILST action. Open strings in the tensionless regime come with some surprises. Working with the Dirichlet analog of the open tensile string and focusing on the mode expansions, we find satisfyingly that the symmetries of the theory give rise to the new algebra discovered below in the context of 2d Carroll BCFTs. 

\paragraph{2d Carroll CFTs.} We begin with a quick recap of Carrollian symmetries. As stated in the introduction, the vanishing speed of light limit of the Poincar\'e algebra leads to the Carroll algebra \cite{SenGupta:1966qer,Leblond65}. Geometrically, this is the isometry algebra of a flat Carrollian manifold. A $d$-dimensional Carrollian manifold is defined in terms of a nowhere-vanishing vector field 
$v^\mu$ and a degenerate symmetric rank-two tensor $h_{\mu\nu}$ with a 1-dimensional kernel spanned by the vector field, $v^{\mu}h_{\mu\nu}=0$. 

For the Carroll analog of Minkowski space, it is convenient to use Cartesian coordinates, in terms of which we have $v = (1, 0, \ldots, 0)$ and $h = \text{diag}(0, 1, \ldots, 1)$. The conformal isometries of this flat Carroll structure are generated by vector fields $\xi^\mu$ that solve the Carroll conformal Killing equations
\begin{equation}
    \mathcal{L}_\xi v^\mu = \lambda\, v^\mu,\qquad\qquad \mathcal{L}_\xi h_{\mu\nu} = - 2 \lambda\, h_{\mu\nu}\,.
\label{eq:angelinajolie}
\end{equation}
Solving \eqref{eq:angelinajolie} in 2d (with coordinates $\tau$ and $\sigma$ so that $v^\tau=1$, $h_{\sigma\sigma}=1$, $v^\sigma=h_{\tau\tau}=h_{\tau\sigma}=0$) yields the Caroll conformal Killing vectors $\xi=[\xi_M(\sigma)+\tau\xi_L^\prime(\sigma)]\partial_\tau+\xi_L(\sigma)\partial_\sigma$ whose Lie brackets generate the (centerless) $d$-dimensional conformal Carroll algebra. In gravity parlance, $\xi_M(\sigma)$ generates ``supertranslations'' and $\xi_L(\sigma)$ ``superrotations''. 

Introducing Fourier modes for these two functions [see Eq.~\eqref{eq:lalapetz} below] and permitting central extensions yields the 2d Carroll conformal algebra
 \begin{subequations}\label{bms3}
  \begin{align}
    \left[L_n,L_m\right]&=(n-m)L_{n+m} +\frac{c_L}{12}(n^3-n)\delta_{n,-m} \\
    \left[L_n,M_m\right]&=(n-m){M}_{n+m}+\frac{c_M}{12}(n^3-n)\delta_{n,-m}\\
    \left[M_n, M_m\right]&=0
\end{align}  
\end{subequations}
where $c_L$ is the Virasoro central charge  and $c_M$ the BMS central charge. 
As mentioned in the introduction, this algebra is isomorphic to BMS$_3$. 

The generators can be understood in terms of the geometry of 3d flat space on its null boundary $\mathscr{I}^\pm$ with cylinder topology $\mathbb{R}\times S^1$, where the $S^1$ sometimes is referred to as ``celestial circle''. The superrotations $L_n$ generate diffeomorphisms of this celestial circle. The supertranslations $M_n$ generate angle-dependent translations along the null direction, i.e.~along advanced or retarded time. In 3d Einstein gravity, an asymptotic symmetry analysis proves that the Virasoro central charge vanishes while the BMS central charge scales inversely to Newton's constant, $c_L=0$ and $c_M=3/G$ \cite{Barnich:2006av}. One can generate a non-zero $c_L$ in topologically massive gravity with asymptotically flat boundary conditions \cite{Bagchi:2012yk}.  

\paragraph{2d Carroll CFTs with boundaries.} We now consider 2d CCFTs with boundaries. We look at these theories defined on the null cylinder with Carroll structure
\begin{equation}
    v^\mu\partial_\mu=\partial_\tau\quad\qquad \extd s^2 = 0\cdot \extd\tau^2 + \extd\sigma^2\quad(\sigma\sim\sigma+2\pi)\,.
\end{equation}
The generators of the 2d CCFT on the cylinder are given by 
\begin{equation}
    L_n=e^{in\sigma}(\partial_\sigma+in\tau\,\partial_\tau)\qquad\qquad M_n=e^{in\sigma}\partial_\tau\,.
    \label{eq:lalapetz}
\end{equation}
Indeed, their Lie bracket algebra produces the centerless version of~\eqref{bms3}~\footnote{%
To get the central terms, one either has to consider quantum operator versions of these generators or canonically realize the symmetries on the state space, which produces central terms in a gravity context \cite{Barnich:2006av,Bagchi:2012yk}. To ease the notation, we shall freely change the meaning of the generators $L_n,M_n$ as either producing the centerless or the full Carroll conformal algebra, and the correct meaning is always clear from the context. 
}.

\begin{figure}[ht]
\centering
    \includegraphics[scale=0.8]{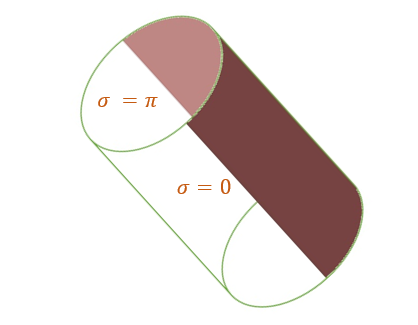}
    \caption{Null cylinder with boundaries}
    \label{fig2}
\end{figure}

We now place boundaries on the cylinder at $\sigma=0,\pi$, see the schematic Fig.~\ref{fig2}. The operators that leave the boundary invariant cannot have a non-zero action of $\partial_\sigma$ at the boundaries. Since the $L_n$ generators have such terms, a na{\"i}ve guess would be to drop all $L_n$. This is, of course, not very interesting as it will leave us with the abelian algebra of the $M_n$. 

We now devise a set of generators that does not have these problematic $\partial_\sigma$ terms at the boundaries, namely
\begin{equation}
    \mathcal{O}_n := L_n-L_{-n} = 2i\sin(n\sigma)\,\partial_\sigma+2in\tau\cos(n\sigma)\,\partial_\tau\,.
\end{equation}
By construction, these operators obey
$\mathcal{O}_n=\pm 2in\tau\,\partial_\tau$ for $\sigma=0,\pi$  
and thus preserve the boundary conditions. 

At this stage, it is less obvious that we need to introduce a similar redefinition in the supertranslation generators,
\begin{equation}
P_n := M_n + M_{-n} = 2 \cos(n\sigma)\, \partial_\tau 
\label{eq:whatever}
\end{equation} 
or, alternatively, $R_n:=M_n-M_{-n}=2\sin(n\sigma)\,\partial_\tau$. The rationale behind these definitions is the invariance (up to possible overall signs) of the generators ${\cal O}_n,P_n,R_n$ under $n\to-n$, which geometrically is achieved upon identifying $\sigma\to-\sigma$. This property allows arriving at the results above through a Carroll limit of relativistic 2d BCFTs.

\paragraph{Algebra of boundary CCFT operators.}
Having defined a set of operators that satisfy the boundary conditions we imposed on the Carrollian null cylinder, we compute the resulting symmetry algebra, finding 
\begin{subequations}\label{bdyCarr}
  \begin{align}
    \left[\mathcal{O}_n,\,\mathcal{O}_m\right]&=(n-m)\mathcal{O}_{n+m}-(n+m)\mathcal{O}_{n-m}\\
    \left[\mathcal{O}_n,\,P_m\right]&=(n-m)P_{n+m}+(n+m)P_{n-m}\nonumber \\  &\quad +\frac{c_M}{12}(n^3-n)(\delta_{n,-m}+\delta_{n,m})\\
    \left[P_n,\,P_m\right]&=0
\end{align}  
\end{subequations}
where $\mathcal{O}_{-n}=-\mathcal{O}_n$ and $P_{-n}=P_n$, $\forall{n}\in\mathbb{Z}$. We call this algebra the Boundary Conformal Carrollian Algebra (BCCA)~\footnote{%
A different basis for the BCCA is given by the generators $Q_n:=\mathcal{O}_n+P_n$ with commutators $[Q_n,Q_m]=(n-m)(\frac32Q_{n+m}+\frac12Q_{-n-m})+(n+m)(\frac12Q_{m-n}-\frac12Q_{n-m})+\frac{c_M}{6}(n^3-n)\delta_{n,-m}$.
}. To the best of our knowledge, this is a new algebra, which opens up innumerable possibilities for further studies, given the wide-ranging applications Carrollian symmetries are finding all over theoretical physics. Further below, we focus on one such application namely open null strings.

Above, we chose specifically the operator \eqref{eq:whatever} in the supertranslation sector. However, we would have obtained the same commutation relations had we chosen $R_n$ or $M_n$ instead of $P_n$. 

\paragraph{Boundary Carroll from contraction.}  We now obtain the Carroll boundary conformal algebra \eqref{bdyCarr} as a novel contraction of a single copy of the Virasoro algebra $[\mathbf{L}_n,\,\mathbf{L}_m]=(n-m)\,\mathbf{L}_n+\frac{c}{12}\,(n^3-n)\,\delta_{n,-m}$, 
\begin{equation}
    \mathcal{O}_n := \mathbf{L}_n - \mathbf{L}_{-n}\qquad P_n := \epsilon \,(\mathbf{L}_n + \mathbf{L}_{-n})\qquad \epsilon \to 0
    \label{eq:contraction}
\end{equation}
with the central charge $c_M=\lim_{\epsilon\to0}(2\epsilon\,c)$. The contraction \eqref{eq:contraction} can be understood as the usual Carrollian contraction of two Virasoros (with generators $\L_n,\bar\L_n$) to BMS$_3$ \cite{Barnich:2006av}, $L_{n}=\L_{n}-\bar{\L}_{-n}$, $M_{n}=\epsilon\,(\L_n + \bar{\L}_{-n})$, after identifying the Virasoro subalgebra $\mathbf{L}_{n}=\L_{n}+\bar{\L}_{n}$ that leaves invariant the boundary. This Virasoro subalgebra is the Lorentzian avatar of the BCCA \eqref{bdyCarr}.

\paragraph{Global subalgebra and central terms.} The BCCA \eqref{bdyCarr} has a global subalgebra given by the generators $\{ P_0, P_1, \mathcal{O}_1\}$. These close to form the 2d Poincar\'e algebra iso$(1,1)$,
\begin{equation}
    [\mathcal{O}_1,P_0] = 2 P_1 \qquad [\mathcal{O}_1,P_1] = 2 P_0\qquad[P_0,P_1]=0\,.
    \label{eq:cghs1}
\end{equation}
As expected, the global subalgebra \eqref{eq:cghs1} is blind to the central charge $c_M$. Perhaps less expected, iso$(1,1)$ comes with its own central extension not present in the BCCA \eqref{bdyCarr}, where the last commutator is replaced by $[P_0,P_1]=Z$, with $[Z,\bullet]=0$, an algebra featuring prominently in the Callan--Giddings--Harvey--Strominger model of 2d dilaton gravity \cite{Callan:1992rs,Cangemi:1992bj}. Centrally extended iso$(1,1)$ is also known as ``Maxwell algebra'', arises in the flat space or near horizon limits of the Jackiw--Teitelboim model, and captures generic aspects of non-extremal horizons, see \cite{Afshar:2019axx,Godet:2020xpk} and Refs.~therein. It should be rewarding to find a similar universal application of our BCCA \eqref{bdyCarr}. The remainder of our Letter is an example of such an application.

\paragraph{Open null strings.} Tensionless or null strings (which are synonymous in flat spacetimes) have been studied since Schild \cite{Schild:1976vq}. However, most of the attention has been devoted to the closed null string. Below, we investigate what happens when we consider open null strings. 

The dynamics of null strings on a flat background is governed by the ILST action \cite{Isberg:1993av}
\begin{equation}
    S_{\text{ILST}} = \frac{1}{4\pi c^\prime}\,\int_{\cal M} \extd^2x\, V^\alpha V^\beta(\partial_\alpha X^\mu)(\partial_\beta X^\nu) \eta_{\mu\nu}
    \label{eq:ILST}
\end{equation}
where $V^\alpha$ is a vector density that essentially plays the role of the (degenerate) inverse worldsheet metric, $X^\mu$ are the target space coordinates, and the indices $\alpha,\beta$ are worldsheet indices. The dimensionful normalization constant $c^\prime$ is arbitrary and will turn out to be useful in our limiting construction below; it is the tensionless relative of the inverse string tension $\alpha^\prime$ in standard string theory, reminiscent of the effective coupling constant ${\hbar}c$ that remains finite in the ``tantum gravity'' limit $\frac1\hbar,c,G_N\to0$ \cite{EFG:2024}. 

Below, we use the worldsheet coordinates $x^\alpha=(\tau,\sigma)$ and assume the 2d manifold $\cal M$ has a boundary $\partial\cal M$ located at $\sigma=0,\pi$, precisely as in Fig.~\ref{fig2}. 

Variation of the ILST action with respect to the fields $X^\mu$ and $V^\alpha$ yields the bulk equations of motion/constraints
\begin{equation}
\partial_\alpha(V^\alpha V^\beta\partial_\beta X^\nu)\eta_{\mu\nu}=0=
V^\beta\eta_{\mu\nu}(\partial_\alpha X^\mu)(\partial_\beta X^\nu)
\label{eq:eom}
\end{equation}
and the boundary requirement
\begin{equation}
\int_{\partial\cal M}\extd\tau\,n_\alpha V^\alpha V^\beta(\partial_\beta X^\nu)\eta_{\mu\nu}\,\delta X^\mu = 0
\label{eq:beom}
\end{equation}
where $n_\alpha$ is normal to the boundary $\partial\cal M$.

The first variation of the ILST action \eqref{eq:ILST} vanishes on-shell for three simple sets of boundary conditions: 
\begin{subequations}
\label{unusual}
\begin{align}
    \delta X^\mu \big|_{\sigma=0,\pi}&=0 \qquad \text{Dirichlet}\label{eq:Dirichlet}\\ 
    V^\beta\partial_\beta X^\nu \big|_{\sigma=0,\pi}&=0 \qquad \text{Neumann}\\
    n_\alpha V^\alpha\big|_{\sigma=0,\pi} &=0 \qquad \text{Null.} \label{eq:DerDritteMann}
    \end{align}
\end{subequations}
The first two conditions above give the usual Dirichlet and Neumann conditions like for the relativistic tensile string. 

The third condition \eqref{eq:DerDritteMann} is specific to the null string. This new boundary condition, which is geometric in nature, raises possible connections to the statement that in the tensionless limit the distinction between closed and open strings blur (see e.g. \cite{Francia:2007qt, Bagchi:2019cay}). Although this is an intriguing option to consider, for the moment we explore the other two options as we are interested in understanding open null strings that emerge as a limit of relativistic tensile open strings. 

\paragraph{Mode expansion of open null strings.} We now focus on Dirichlet boundary conditions \eqref{eq:Dirichlet}. We work in the gauge $V^\alpha=(1,0)$. The constraints/bulk equations of motion \eqref{eq:eom} simplify to
\begin{equation}
\ddot{X}^\mu=0\qquad\quad  \dot{X}^2=0\qquad\quad \dot{X}\cdot X'=0
 \label{v0eom}
\end{equation}
where dots (primes) denote derivatives with respect to the worldsheet null time $\tau$ (the spatial coordinate $\sigma$) and the inner products are taken with the target space Minkowski metric, e.g., $\dot{X}\cdot X'=\dot{X}^\mu (X')^\nu \eta_{\mu\nu}$.

With fixed boundary values $X^{\mu} (0)= x_{0}^{\mu}$, $X^{\mu} (\pi)=x_{1}^{\mu}$, the most general solution to the equations of motion $\ddot X^\mu=0$ is
\begin{equation}
X^{\mu} (\tau ,\sigma) = x_{0}^{\mu} + \frac{x_{1}^{\mu}-x_{0}^{\mu}}{\pi}\,\sigma + a^{\mu} (\sigma) + \tau \,b^{\mu} (\sigma)\, ,
\end{equation}
where $a^{\mu} (\sigma)$ and $b^{\mu} (\sigma)$ are real functions that vanish at the boundary $\sigma=0,\pi$. Expanding them in modes $\sin(n\sigma)$, we can write the solution as
\begin{align}
    X^{\mu}(\tau,\sigma)=x_0^{\mu}+\sqrt{2c'}&C^{\mu}_{0}\sigma +i\sqrt{\frac{c'}{2}}\sum_{n\neq0}\frac{1}{n}\big[(C^{\mu}_{n}+C^{\mu}_{-n})\nonumber \\ &-in\tau(C^{\mu}_{n}-C^{\mu}_{-n})\big]e^{-in\sigma}\label{modeexpansion}
\end{align}
where reality enforces $(C_{n}^\mu)^{\dagger}=C_{-n}^\mu$ and we defined $C_0^\mu=(x_1^\mu-x_0^\mu)/(\sqrt{2c^\prime}\pi)$. The mode expansion \eqref{modeexpansion} of the tensionless bosonic open string by construction is consistent with Dirichlet boundary condition \eqref{eq:Dirichlet}. 

\paragraph{Constraint analysis and symmetry algebra.} Having solved the bulk and boundary equations of motion for tensionless open strings, we now address the bulk constraints $\dot{X}^2=\dot{X}\cdot{X}^\prime=0$. From the mode expansion \eqref{modeexpansion}, we calculate $\dot X^\mu, X'^\mu$ and the constraint equations can be recast in terms of the bilinear generators
\begin{subequations}\label{eq:con}
\begin{align}
&\dot{X}^2=\sum_{m} P_m e^{-im\sigma} = 0, \\
& \dot{X}\cdot{X}^\prime = \sum_{m}\big(\mathcal{O}_m-im\tau P_m\big)e^{-im\sigma}=0\,, 
\end{align}
\end{subequations}
where we have defined
\begin{subequations}
    \label{eq:bi}
\begin{align}
 P_n&:=\sum_{m}\frac{1}{4}(C_m^\mu-C_{-m}^\mu)(C_{n-m}^\nu-C_{m-n}^\nu)\eta_{\mu\nu},\\
\mathcal{O}_{m}&:=\sum_{n}\frac{1}{2}(C_n^\mu-C_{-n}^\mu)(C_{m-n}^\nu+C_{n-m}^\nu)\eta_{\mu\nu}.
\end{align}
\end{subequations}
%We are considering the classical theory. Looking forward, when quantising null open string, as usual, one should not enforce the constraints \eqref{eq:con} as strong operator identities but as conditions acting on physical states; 
In a potential construction of a quantum theory of null strings, the classical constraints~\eqref{eq:con} should, as usual, not be enforced as strong operator identities but as conditions acting on physical states;
see \cite{Bagchi:2020fpr} for more on these conditions and the associated physical vacua for the closed null string.

Canonical commutations $[X^\mu(\tau,\sigma),\,\dot{X}^\nu(\tau,\sigma^\prime)]=2{\pi}ic^\prime\eta^{\mu\nu}\delta(\sigma-\sigma^\prime)$, requires
\begin{equation}
[C_n^\nu,\,C_m^\mu] = n\,\delta_{n,-m}\,\eta^{\mu\nu}\,.
\label{eq:osc}
\end{equation}
By virtue of the commutation relations \eqref{eq:osc}, the algebra of the bilinear constraint generators $\mathcal{O}_m$ and $P_m$ is precisely the BCCA \eqref{bdyCarr} (with $c_M=0$). Central terms would be generated in the quantum theory. Thus, we have found an explicit realization of the BCCA that we constructed previously using purely field theoretic arguments. The open null string is, therefore, an example of a physical system where the BCCA emerges; as we showed above, its role is analogous to the boundary Virasoro algebra in the context of tensile open strings.

\paragraph{Null open string from tensile open string.} The mode expansion for the (bosonic) tensile open string (see, e.g., \cite{Zwiebach:2004tj})
\begin{multline}
\label{opentensilemode}
     X^{\mu}(\tau,\sigma)=x^{\mu}_0+\sqrt{2\alpha'}\alpha_0^\mu\sigma+i\sqrt{\frac{\alpha'}{2}}\sum_{n\neq0}\frac{1}{n}\cdot\\
     \cdot\Big[\alpha^{\mu}_{n} e^{-in(\tau+\sigma)} + \alpha^{\mu}_{-n}e^{in(\tau-\sigma)}\Big]
\end{multline}
permits taking the tensionless limit by singularly rescaling the worldsheet time coordinate
\begin{equation}
\label{contraction}
    \tau\to\epsilon\tau\qquad\sigma\to\sigma \quad \text{and} \quad\alpha'\to c'/\epsilon\qquad\epsilon\to 0\,.
\end{equation}
Here, $\alpha^\prime$ is the inverse string tension. Comparing the mode expansion \eqref{opentensilemode} for small $\epsilon$ with the open null string mode expansion \eqref{modeexpansion} reveals that they are identical provided we relate the corresponding oscillators by the Bogoliubov transformation
\begin{equation}
\label{openbv}
C^{\mu}_{n}=\frac{1}{2}\Big(\sqrt{\epsilon}+\frac{1}{\sqrt{\epsilon}}\Big)\,\alpha^{\mu}_{n}-\frac{1}{2}\Big(\sqrt{\epsilon}-\frac{1}{\sqrt{\epsilon}}\Big)\,{\alpha}^{\mu}_{-n}
\end{equation}
that preserves the Poisson structure. Like in the closed string construction \cite{Bagchi:2015nca}, the Bogoliubov transformation \eqref{openbv} becomes singular in the limit $\epsilon\to 0$.

Inserting the Bogoliubov transformation \eqref{openbv} into the expressions \eqref{eq:bi} for the bilinears yields expressions
\begin{subequations}\label{bccg}
    \begin{align}
     \mathcal{O}_n &=\frac{1}{2}\sum_{p}(\alpha_p\cdot\alpha_{n-p}-\alpha_{-p}\cdot\alpha_{p-n})\\
        P_n &=\frac{\epsilon}{2}\sum_{p} \left(\alpha_p\cdot\alpha_{n-p}+\alpha_{-p}\cdot\alpha_{p-n}\right)\label{bccg1}
        \end{align}
\end{subequations}
in terms of the tensile oscillator modes that appear in the Virasoro generators $\textbf{L}_n=\tfrac{1}{2}\sum_{p}\alpha_p\cdot\alpha_{n-p}$. Consistently, by expressing \eqref{bccg} in terms of the Virasoro generators $\textbf{L}_n$ we recover the relations \eqref{eq:contraction}. So, from the limiting analysis of the string above we also retrieve the BCCA \eqref{bdyCarr}.

\paragraph{From closed to open null strings.} The tensile open string with Dirichlet boundary conditions can be obtained from closed tensile strings by an identification of the two independent oscillators of the closed string, viz. $\alpha_n = -\tilde{\alpha}_n$. The closed null string can also be written down in terms of two sets of oscillators $C, \tilde{C}$ that obey commuting harmonic oscillator algebras. In this basis, the solution to the equations of motion \cite{Bagchi:2015nca, Bagchi:2020fpr}
\begin{multline}
\label{tlessmode}
    X^{\mu}(\tau,\sigma)=x^{\mu}+\sqrt{\tfrac{c'}{2}}\,(C^{\mu}_{0}-\Tilde{C}^{\mu}_{0})\,\sigma+\sqrt{\tfrac{c'}{2}}\,(C^{\mu}_{0}+\Tilde{C}^{\mu}_{0})\,\tau\\+i\sqrt{\tfrac{c'}{2}}\sum_{n\neq0}\tfrac{e^{-in\sigma}}{n}[(C^{\mu}_{n}-\Tilde{C}^{\mu}_{-n})-in\tau(C^{\mu}_{n}+\Tilde{C}^{\mu}_{-n})]
\end{multline}
recovers the open null string \eqref{modeexpansion} by an identification mirroring the tensile case:
\begin{equation}
 C^\mu_n = - \tilde{C}^\mu_n\,.   
\end{equation}
This procedure is another way to obtain open null strings and provides an independent cross-check for our results above. 

\paragraph{Conclusions and outlook.} In this Letter, we have initiated a study of Carrollian CFTs with boundaries and provided an explicit realization in terms of open null strings. This work should engender research into a whole range of new topics, including applications of BCCFTs to various potential condensed matter systems like fractons and systems with flat bands that contain boundaries. 

The fact that Carrollian CFTs are at the heart of flat holography means that a flat version of the AdS/BCFT correspondence would involve the symmetries we have unearthed in this Letter. 

Our definition \eqref{eq:whatever} of supertranslation generators compatible with the presence of a boundary generalizes straightforwardly to supertranslation generators with arbitrary spin $s$ \cite{Grumiller:2019fmp,Campoleoni:2020ejn,Campoleoni:2021blr,Figueroa-OFarrill:2024wgs}, $P_n^{(s)}:=M_n^{(s)}+M_{-n}^{(s)}$. In our Letter, we studied the case $s=2$. For general values of $s$, the only change in the BCCA \eqref{bdyCarr} is the mixed commutator,
\begin{align}
\big[\mathcal{O}_n,\,P_m^{(s)}\big]&=\big((s-1)n-m\big)\,P_{n+m}^{(s)} \nonumber\\
&\, + \big((s-1)n+m\big)\,P_{n-m}^{(s)} + \textrm{central}
\end{align}
where the form of the central terms depends on the choice of $s$. Prominent examples for choices of $s$ (besides $s=2$) are $s=0$ (BMS$_2$) \cite{Afshar:2019axx,Afshar:2021qvi} and $s=1$ (twisted warped Witt) \cite{Detournay:2012pc,Afshar:2015wjm}. Given the various applications for $s\neq 2$, it would be gratifying to apply our methods and generalize our results to arbitrary spin $s$.

Our construction of classical open null strings provides the first step to their quantum description, an outstanding open issue. The quantization of the closed null string led to three different quantum null strings \cite{Bagchi:2020fpr}. It would be of great interest to see how this fits into the new symmetry structure of the open null string, given by our new algebra \eqref{bdyCarr}. The construction of null D-branes should naturally follow. The extension to null open superstrings is another immediate direction, which is currently under investigation. (See \cite{Bagchi:2016yyf, Bagchi:2017cte} for discussions on closed null superstrings.)

Finally, we have only considered the Dirichlet boundary condition in \eqref{unusual}. It is natural to investigate the other two boundary conditions; especially intriguing is the third option \eqref{eq:DerDritteMann}, given its uniqueness to the null string without tensile precedent. We intend to return to these and other research directions soon. 

%%%%%%%%%%%%%%%%%%%%%%%
%%% ACKNOWLEDGMENTS %%%
%%%%%%%%%%%%%%%%%%%%%%%

\bigskip

\noindent \textbf{Acknowledgements.} We thank Aritra Banerjee, Ritankar Chatterjee, and Shahin Sheikh-Jabbari for useful discussions. 

AB is partially supported by a Swarnajayanti Fellowship from the Science and Engineering Research Board
(SERB) under grant SB/SJF/2019-20/08. AB acknowledges support from a Research in Teams project by the Erwin Schr\"odinger International Institute for Mathematics and Physics (ESI) in June/July 2022, where this project commenced. AB and SC acknowledge ANRF grant CRG/2022/006165. DG was supported by the Austrian Science Fund (FWF), projects P 32581, P 33789, and P 36619 \href{https://www.fwf.ac.at/forschungsradar/10.55776/P36619}{[grant DOI:10.55776/P36619]}. DG acknowledges support by the OeAD travel grant IN 04/2022 and thanks Rudranil Basu for hosting them at BITS Pilani in Goa in February 2024 through the grant DST/IC/Austria/P-9/202 (G). 
AB, SF, and DG acknowledge the hospitality of the ESI during the Thematic Programme: ``Carrollian Physics and Holography'' in April 2024. SC, PP and PC would like to acknowledge the organisers of the workshop ``Holography, Strings and Other Fun things" organised at BITS Pilani Goa in February 2024. SC and PP would also like to thank the hospitality of IISER Pune during the conference ``Future Perspectives on QFT and Strings'' in July 2024. PP is supported by an IIT Kanpur Institute Assistantship for Postdoctoral Fellow.

%%%%%%%%%%%%%%%%%%
%%% REFERENCES %%%
%%%%%%%%%%%%%%%%%%

%\bibliographystyle{utphys2}
%\bibliography{ccft}
\providecommand{\href}[2]{#2}\begingroup\raggedright\endgroup

\end{document}